\documentclass{article}
\usepackage{amssymb}
\usepackage{amsfonts}
\usepackage{amsmath}
\usepackage{harvard}

\input{tcilatex}

\begin{document}

\title{\textbf{Symmetries and Hamiltonian formalism for complex materials}}
\author{$%
{{}^\circ}%
$\textbf{Gianfranco Capriz} and *\textbf{Paolo Maria Mariano} \\
$%
{{}^\circ}%
$Dipartimento di Matematica,\\
Universit\`{a} di Pisa,\\
via Buonarroti 2, I-56127 Pisa (Italy)\\
e-mail: capriz@dm.unipi.it ;\\
*Dipartimento di Ingegneria Strutturale e Geotecnica,\\
Universit\`{a} di Roma "La Sapienza",\\
via Eudossiana 18, I-00184 Roma (Italy)\\
e-mail: paolo.mariano@uniroma1.it}
\maketitle

\begin{abstract}
Preliminary results toward the analysis of the Hamiltonian structure of
multifield theories describing complex materials are mustered: we involve
the invariance under the action of a general Lie group of the balance of
substructural interactions. Poisson brackets are also introduced in the
material representation to account for general material substructures. A
Hamilton-Jacobi equation suitable for multifield models is presented.
Finally, a spatial version of all these topics is discussed without making
use of the notion of paragon setting.
\end{abstract}

\section{Lagrangian and Hamiltonian descriptions of elastic bodies with
substructure}

In standard continuum mechanics, each material element of a body is
\textquotedblleft collapsed\textquotedblright\ into the place occupied by
its centre of mass; let $\mathbf{X}$ be that place in the reference
placement; the set of all $\mathbf{X}$ is taken to be a fit region $\mathcal{%
B}_{0}$\ of the Euclidean space.

Sometimes such simplicistic model of physical reality is insufficient; then,
to render the picture adequate, the material element must be portrayed as a 
\emph{system} and at least some coarse grained descriptor $\mathbf{\nu }$
(an order parameter) enters the picture.

Here, as in [1] (see for other details and additional results [2-5]), we
take $\mathbf{\nu }$ as an element of some differentiable manifold $\mathcal{%
M}$, and presume that physical circumstances impose a single choice of
metric and of connection for $\mathcal{M}$.

We also assume that the region occupied by the body in the current placement
be obtained through a sufficiently smooth mapping $\mathbf{x:}\mathcal{B}%
_{0}\rightarrow \mathcal{E}$ ; so that the current place of a material
element at $\mathbf{X}$ in $\mathcal{B}_{0}$ is given by $\mathbf{x}\left( 
\mathbf{X}\right) $; and $\mathcal{B}=\mathbf{x}\left( \mathcal{B}%
_{0}\right) $ is also \emph{fit}; we denote as usual with $\mathbf{F}$ the
placement gradient. We presume also that another sufficiently smooth mapping 
$\mathbf{\nu :}\mathcal{B}_{0}\rightarrow \mathcal{M}$ shows the present
value of the order parameter at $\mathbf{X}$. A \emph{motion} is a pair of
time-parametrized families $\mathbf{x}_{t}\left( \mathbf{X}\right) =\mathbf{x%
}\left( \mathbf{X},t\right) $ and $\mathbf{\nu }_{t}\left( \mathbf{X}\right)
=\mathbf{\nu }\left( \mathbf{X},t\right) $, twice differentiable with
respect to time. Rates in the material representation are indicated with $%
\mathbf{\dot{x}}\left( \mathbf{X},t\right) $ and $\mathbf{\dot{\nu}}\left( 
\mathbf{X},t\right) $.

We restrict here our attention to bodies for which a Lagrangian density $%
\mathcal{L}$\ exists, so that the total Lagrangian $L$ of the body is given
by 
\begin{equation}
L_{\mathcal{B}_{0}}=\int_{\mathcal{B}_{0}}\mathcal{L}\left( \mathbf{X,x,\dot{%
x},F,\nu ,\dot{\nu},}\nabla \mathbf{\nu }\right) .  \label{1}
\end{equation}%
the covariant gradient $\nabla \mathbf{\nu }$\ being based on the mandatory
connection. We presume that $\mathcal{L}$ be of the form%
\begin{equation}
\mathcal{L}\left( \mathbf{X,x,\dot{x},F,\nu ,\dot{\nu},}\nabla \mathbf{\nu }%
\right) =\frac{1}{2}\rho _{0}\left\Vert \mathbf{\dot{x}}\right\Vert
^{2}+\rho _{0}\chi \left( \mathbf{\nu ,\dot{\nu}}\right) -\rho _{0}e\left( 
\mathbf{X,F,\nu ,}\nabla \mathbf{\nu }\right) -\rho _{0}w\left( \mathbf{%
x,\nu }\right) ,  \label{2}
\end{equation}%
where $\rho _{0}$\ is the referential mass density (conserved during the
motion), $\chi $\ the \emph{kinetic co-energy} (see [1], p. 19) associated
with the substructure, $e$ the elastic energy density and $w$\ the density
of the potential of external actions, all per unit mass. Below we use the
notation $\mathbf{b=-\partial }_{\mathbf{x}}w$ for the density standard
external actions and $\mathbf{\beta =-\partial _{\mathbf{\nu }}}w$\ for the
substructural ones. The kinetic energy density $\rho _{0}\kappa \left( 
\mathbf{\nu ,\dot{\nu}}\right) $\ pertaining to the substructure is the
partial Legendre transform of $\chi $ with respect to $\mathbf{\dot{\nu}}$.

If $\mathcal{L}$ is sufficiently smooth, we may apply standard procedures to
derive Lagrange equations for the functional $L_{\mathcal{B}_{0}}$:%
\begin{equation}
\overset{\cdot }{\overline{\partial _{\mathbf{\dot{x}}}\mathcal{L}}}%
=\partial _{\mathbf{x}}\mathcal{L}-Div\partial _{\mathbf{F}}\mathcal{L},
\label{3}
\end{equation}%
\begin{equation}
\overset{\cdot }{\overline{\partial _{\mathbf{\dot{\nu}}}\mathcal{L}}}%
=\partial _{\mathbf{\nu }}\mathcal{L}-Div\partial _{\nabla \mathbf{\nu }}%
\mathcal{L}.  \label{4}
\end{equation}%
where $Div$ is the divergence calculated with respect to $\mathbf{X}$, i.e. $%
Div=tr\nabla $.

Put%
\begin{equation}
\mathfrak{H}=\mathbf{\dot{x}\cdot }\partial _{\mathbf{\dot{x}}}\mathcal{L}+%
\mathbf{\dot{\nu}\cdot }\partial _{\mathbf{\dot{\nu}}}\mathcal{L}-\mathcal{L}%
.  \label{5}
\end{equation}%
Clearly, $\mathfrak{H}$\ is the density of the total energy. In fact, since $%
\partial _{\mathbf{\dot{\nu}}}\mathcal{L}=\rho _{0}\partial _{\mathbf{\dot{%
\nu}}}\chi $, the term $\mathbf{\dot{\nu}}\cdot \partial _{\mathbf{\dot{\nu}}%
}\chi -\chi $ in (\ref{5}) coincides with the substructural kinetic energy
density $\kappa \left( \mathbf{\nu ,\dot{\nu}}\right) $ (hence the presence
of $\chi \mathbf{\ }$rather than $\kappa $ in the expression of $\mathcal{L}$%
), then%
\begin{equation}
\mathfrak{H}=\frac{1}{2}\rho _{0}\left\Vert \mathbf{\dot{x}}\right\Vert
^{2}+\rho _{0}\kappa \left( \mathbf{\nu ,\dot{\nu}}\right) +\rho _{0}e\left( 
\mathbf{X,F,\nu ,}\nabla \mathbf{\nu }\right) +\rho _{0}w\left( \mathbf{x},%
\mathbf{\nu }\right) ,  \label{6}
\end{equation}%
as asserted.

The \emph{balance of energy} can be expressed in terms of $\mathfrak{H}$ as
follows%
\begin{equation}
\mathfrak{\dot{H}}-Div\left( \mathbf{\dot{x}P+\dot{\nu}}\mathcal{S}\right)
=0,  \label{7}
\end{equation}%
where $\mathbf{P}$ and $\mathcal{S}$ are respectively the \emph{%
Piola-Kirchhoff stress} and the referential \emph{microstress}%
\begin{equation}
\mathbf{P=-}\partial _{\mathbf{F}}\mathcal{L}\text{ \ \ , \ \ }\mathcal{S}%
=-\partial _{\nabla \mathbf{\nu }}\mathcal{L}.  \label{8}
\end{equation}

That (\ref{7}) is true follows from direct computation.

Equations (\ref{3}) and (\ref{4}) lead us to an appropriate version of N\"{o}%
ther theorem (see [4] p. 29); here we follow the program of [6] (p. 284). We
consider some virtual motion of our system, by assigning two one-parameter
families $\mathbf{f}_{s_{i}}^{i}$ of sufficiently smooth point valued
diffeomorphisms, $i=1,2$, acting respectively on $\mathcal{B}_{0}$ and $%
\mathcal{E}$, and a Lie group $G$ of transformations of $\mathcal{M}$. We
indicate with a prime the derivative with respect to the relevant $s$.

\begin{enumerate}
\item At each $s_{1}$, $\mathbf{f}_{s_{1}}^{1}$ acts on $\mathcal{B}_{0}$\
so that $\mathbf{X\longmapsto f}_{s_{1}}^{1}\left( \mathbf{X}\right) \in 
\mathcal{E}$, and is isocoric (no virtual change of density), i.e. $Div%
\mathbf{f}_{s_{1}}^{1\prime }=0$; $\mathbf{f}_{0}^{1}$ is the identity. We
put $\mathbf{f}_{0}^{1\prime }\left( \mathbf{X}\right) =\mathbf{w}\left( 
\mathbf{X}\right) $, then $\mathbf{w}^{\prime }\mathbf{=0}$.

\item At each $s_{2}$, $\mathbf{f}_{s_{2}}^{2}$ is a diffeomorphism that
transforms $\mathcal{E}$ into itself. We assume that $\mathbf{f}_{0}^{2}$ is
the identity and put $\mathbf{f}_{0}^{2\prime }\left( \mathbf{x}\right) =%
\mathbf{v}\left( \mathbf{x}\right) $.

\item A Lie group $G$, containing $SO\left( 3\right) $, acts on $\mathcal{M}$
and the infinitesimal generator of this action at $\mathbf{\nu }$ is
indicated with $\xi _{\mathcal{M}}\left( \mathbf{\nu }\right) $ (see [7], p.
256); $\mathbf{\nu }_{g}$ is the value of $\mathbf{\nu }$ after the action
of $g\in G$. If we consider a one-parameter trajectory $s_{3}\mapsto
g_{s_{3}}\in G$ such that $g_{0}$ is the identity, we have also $%
s_{3}\mapsto \mathbf{\nu }_{g_{s_{3}}}$ and $\xi _{\mathcal{M}}\left( 
\mathbf{\nu }\right) =\frac{d}{ds_{3}}\mathbf{\nu }_{g_{s_{3}}}\left\vert
_{s_{3}=0}\right. $. When $G$ coincides with the special orthogonal group $%
SO\left( 3\right) $, we identify $\xi _{\mathcal{M}}\left( \mathbf{\nu }%
\right) $ with $\mathcal{A}\mathbf{\dot{q}}$, being $\mathbf{\dot{q}}$ the
characteristic vector of a rotational rigid velocity and $\mathcal{A}$ a
linear operator mapping vectors into elements of the tangent space of $%
\mathcal{M}$, namely, if $\mathbf{\nu }_{\mathbf{q}}$ is the value of the
order parameter measured by an observer after a rotation $\mathbf{q}$, then $%
\mathcal{A}=\frac{d\mathbf{\nu }_{\mathbf{q}}}{d\mathbf{q}}\left\vert _{%
\mathbf{q}=\mathbf{0}}\right. $.
\end{enumerate}

Henceforth, to simplify notations, we use $\mathbf{f}^{1}$, $\mathbf{f}^{2}$
and $\mathbf{\nu }_{g}$\ to indicate $\mathbf{f}_{s_{1}}^{1}\left( \mathbf{X}%
\right) $, $\mathbf{f}_{s_{2}}^{2}\left( \mathbf{x}\right) $, $\mathbf{\nu }%
_{g_{s_{3}}}\left( \mathbf{X}\right) $, and write $\left\vert _{0}\right. $
for $\left\vert _{s_{1}=0,s_{2}=0,s_{3}=0}\right. $. Moreover, $grad$
indicates the gradient with respect to $\mathbf{x}$.

We say that $\mathcal{L}$ is \emph{invariant} with respect to $\mathbf{f}%
^{i} $'s and $G$ when%
\begin{equation*}
\mathcal{L}\left( \mathbf{X,x,\dot{x},F,\nu ,\dot{\nu},}\nabla \mathbf{\nu }%
\right) =
\end{equation*}%
\begin{equation}
=\mathcal{L}\left( \mathbf{f}^{1}\mathbf{,f}^{2}\mathbf{,}\left( grad\mathbf{%
\mathbf{f}}^{2}\right) \mathbf{\dot{x},}\left( grad\mathbf{\mathbf{f}}%
^{2}\right) \mathbf{F}\left( \nabla \mathbf{f}^{1}\right) ^{-1}\mathbf{,%
\mathbf{\nu }_{g},\mathbf{\dot{\nu}}_{g},}\left( \nabla \mathbf{\nu }%
_{g}\right) \left( \nabla \mathbf{f}^{1}\right) ^{-1}\right) .  \label{11}
\end{equation}

Let us define%
\begin{equation}
\mathcal{Q}=\partial _{\mathbf{\dot{x}}}\mathcal{L\cdot }\left( \mathbf{v}-%
\mathbf{Fw}\right) +\partial _{\mathbf{\dot{\nu}}}\mathcal{L\cdot }\left(
\xi _{\mathcal{M}}\left( \mathbf{\nu }\right) \mathbf{-}\left( \nabla 
\mathbf{\nu }\right) \mathbf{w}\right)  \label{12}
\end{equation}%
\begin{equation}
\mathfrak{F}=\mathcal{L}\mathbf{w+}\left( \partial _{\mathbf{F}}\mathcal{L}%
\right) ^{T}\left( \mathbf{v}-\mathbf{Fw}\right) +\left( \partial _{\nabla 
\mathbf{\nu }}\mathcal{L}\right) ^{T}\left( \xi _{\mathcal{M}}\left( \mathbf{%
\nu }\right) \mathbf{-}\left( \nabla \mathbf{\nu }\right) \mathbf{w}\right) ,
\label{13}
\end{equation}%
where $\mathbf{v}$, $\mathbf{w}$ and $\xi _{\mathcal{M}}\left( \mathbf{\nu }%
\right) $ are as mentioned in items 1, 2, 3.

\textbf{Theorem 1} (N\"{o}ther-like theorem for complex materials). \emph{If
the Lagrangian density} $\mathcal{L}$ \emph{is invariant under} $\mathbf{f}%
_{s_{1}}^{1}$, $\mathbf{f}_{s_{2}}^{2}$\emph{\ and }$\emph{G}$\emph{, then}%
\begin{equation}
\mathcal{\dot{Q}}+Div\mathfrak{F}=0\text{.}  \label{14}
\end{equation}

\emph{Proof}. To prove the theorem, as a first step we note that (\ref{11})
implies%
\begin{equation}
\frac{d}{ds_{1}}\mathcal{L}\left( \mathbf{f}^{1}\mathbf{,f}^{2}\mathbf{,}%
\left( grad\mathbf{\mathbf{f}}^{2}\right) \mathbf{\dot{x},}\left( grad%
\mathbf{\mathbf{f}}^{2}\right) \mathbf{F}\left( \nabla \mathbf{f}^{1}\right)
^{-1}\mathbf{,\nu }_{g},\mathbf{\dot{\nu}}_{g}\mathbf{,}\left( \nabla 
\mathbf{\nu }_{g}\right) \left( \nabla \mathbf{f}^{1}\right) ^{-1}\right)
\left\vert _{0}\right. =0,  \label{15}
\end{equation}%
\begin{equation}
\frac{d}{ds_{2}}\mathcal{L}\left( \mathbf{f}^{1}\mathbf{,f}^{2}\mathbf{,}%
\left( grad\mathbf{\mathbf{f}}^{2}\right) \mathbf{\dot{x},}\left( grad%
\mathbf{\mathbf{f}}^{2}\right) \mathbf{F}\left( \nabla \mathbf{f}^{1}\right)
^{-1}\mathbf{,\nu }_{g},\mathbf{\dot{\nu}}_{g},\left( \nabla \mathbf{\nu }%
_{g}\right) \left( \nabla \mathbf{f}^{1}\right) ^{-1}\right) \left\vert
_{0}\right. =0,  \label{15bis}
\end{equation}%
\begin{equation}
\frac{d}{ds_{3}}\mathcal{L}\left( \mathbf{f}^{1}\mathbf{,f}^{2}\mathbf{,}%
\left( grad\mathbf{\mathbf{f}}^{2}\right) \mathbf{\dot{x},}\left( grad%
\mathbf{\mathbf{f}}^{2}\right) \mathbf{F}\left( \nabla \mathbf{f}^{1}\right)
^{-1}\mathbf{,\nu }_{g},\mathbf{\dot{\nu}}_{g}\mathbf{,}\left( \nabla 
\mathbf{\nu }_{g}\right) \left( \nabla \mathbf{f}^{1}\right) ^{-1}\right)
\left\vert _{0}\right. =0,  \label{15ter}
\end{equation}%
which lead to%
\begin{equation}
\partial _{\mathbf{X}}\mathcal{L}\cdot \mathbf{w-\partial _{\mathbf{F}}}%
\mathcal{L}\mathbf{\cdot }\left( \mathbf{F}\nabla \mathbf{w}\right) \mathbf{%
-\partial _{\nabla \mathbf{\nu }}\mathcal{L}\cdot }\left( \nabla \mathbf{%
\mathbf{\nu }}\nabla \mathbf{\mathbf{w}}\right) \mathbf{=}0,  \label{16}
\end{equation}%
\begin{equation}
\mathbf{\partial _{\mathbf{x}}}\mathcal{L}\cdot \mathbf{v}+\mathbf{\partial
_{\mathbf{\dot{x}}}}\mathcal{L}\mathbf{\cdot }\left( \left( grad\mathbf{v}%
\right) \mathbf{\dot{x}}\right) \mathbf{+\partial _{\mathbf{F}}}\mathcal{L}%
\mathbf{\cdot }\left( \left( grad\mathbf{v}\right) \mathbf{F}\right) \mathbf{%
=}0,  \label{16bis}
\end{equation}%
\begin{equation}
\partial _{\mathbf{\nu }}\mathcal{L}\cdot \xi _{\mathcal{M}}\left( \mathbf{%
\nu }\right) +\partial _{\mathbf{\dot{\nu}}}\mathcal{L}\cdot \xi _{\mathcal{M%
}}^{\prime }\left( \mathbf{\nu }\right) +\partial _{\nabla \mathbf{\nu }}%
\mathcal{L}\cdot \nabla \xi _{\mathcal{M}}\left( \mathbf{\nu }\right) =0,
\label{16ter}
\end{equation}%
as a consequence of the properties listed under items 1, 2 and 3 above.
Then, we calculate the time rate of the scalar $\mathcal{Q}$, the divergence
of the vector $\mathfrak{F}$ and, by using the equations (\ref{3}) and (\ref%
{4}), identifying $s_{3}$\ with $t$, we recognize that%
\begin{equation}
\mathcal{\dot{Q}}+Div\mathfrak{F}=\frac{d}{ds_{1}}\mathcal{L}\left\vert
_{0}\right. +\frac{d}{ds_{2}}\mathcal{L}\left\vert _{0}\right. +\frac{d}{%
ds_{3}}\mathcal{L}\left\vert _{0}\right. ,  \label{17}
\end{equation}%
which proves the theorem.$\square $

\textbf{Remark 1}. As a first special case, we require that $\mathbf{\mathbf{%
f}}^{2}$ alone acts on $\mathcal{L}$ leaving $\mathbf{v}$ \emph{arbitrary}.
By using (\ref{16bis}) we obtain from (\ref{14})%
\begin{equation}
\left( \frac{\partial }{\partial t}\mathbf{\partial _{\mathbf{\dot{x}}}}%
\mathcal{L}-\partial _{\mathbf{x}}\mathcal{L}+Div\partial _{\mathbf{F}}%
\mathcal{L}\right) \cdot \mathbf{v}=\mathbf{0},  \label{18}
\end{equation}%
i.e.%
\begin{equation}
\rho _{0}\mathbf{\ddot{x}=}\rho _{0}\mathbf{b+}Div\mathbf{P},  \label{19}
\end{equation}%
which is the standard equation of balance of momentum.

\textbf{Remark 2}. With $G$ \emph{arbitrary}, we consider its action alone
on $\mathcal{L}$; by using (\ref{16ter}), with the identification $s_{3}=t$,
we obtain from (\ref{14}) that%
\begin{equation}
\left( \frac{\partial }{\partial t}\partial _{\mathbf{\dot{\nu}}}\mathcal{L}%
-\partial _{\mathbf{\nu }}\mathcal{L}+Div\partial _{\nabla \mathbf{\nu }}%
\mathcal{L}\right) \cdot \xi _{\mathcal{M}}\left( \mathbf{\nu }\right) =0
\label{20}
\end{equation}%
or%
\begin{equation}
\rho _{0}\left( \overset{\cdot }{\overline{\partial _{\mathbf{\dot{\nu}}%
}\chi }}-\partial _{\mathbf{\nu }}\chi \right) +\mathbf{z}-\rho _{0}\mathbf{%
\beta }-Div\mathcal{S}=0.  \label{21}
\end{equation}%
$\mathbf{z=-}\rho _{0}\partial _{\mathbf{\nu }}e$ is called \emph{self-force}
in the terminology of [1]. This result assures the \emph{covariance} of the
balance of substructural interactions. When $G$ coincides with $SO\left(
3\right) $, the co-vector into parentheses in (\ref{20}), namely the term
multiplying $\xi _{\mathcal{M}}\left( \mathbf{\nu }\right) $, must be an
element of the null space of $\mathcal{A}^{T}$ (see for details of this
special case [1-4]).

\textbf{Remark 3}. As a second special choice, let $\mathbf{\mathbf{f}}^{2}$
be such that $\mathbf{v=\dot{q}}\times \left( \mathbf{x}-\mathbf{x}%
_{0}\right) $ (with $\mathbf{\dot{q}}$ a rigid rotational velocity $\mathbf{x%
}_{0}$ a fixed point in space) and $G=SO\left( 3\right) $; $\mathbf{\dot{q}}%
\times $ is an element of its Lie algebra, thus $\xi _{\mathcal{M}}\left( 
\mathbf{\nu }\right) =\mathcal{A}\mathbf{\dot{q}}$. If $\mathcal{L}$ is
independent of $\mathbf{x}$ and we assume that only $\mathbf{\mathbf{f}}^{2}$
and $G$ (in the form just defined) act on $\mathcal{L}$, we have%
\begin{equation}
skw\left( \partial _{\mathbf{F}}\mathcal{L}\mathbf{F}^{T}\right) =\mathsf{e}%
\left( \mathcal{A}^{T}\partial _{\mathbf{\nu }}\mathcal{L}+\left( \nabla 
\mathcal{A}^{T}\right) ^{t}\partial _{\nabla \mathbf{\nu }}\mathcal{L}%
\right) \mathbf{,}  \label{22}
\end{equation}%
where $\mathsf{e}$ is Ricci's alternating tensor and $skw\left( \cdot
\right) $ extracts the skew-symmetric part of its argument.

\textbf{Remark 4}. If we require that $\mathbf{f}^{1}$ alone acts on $%
\mathcal{L}$, with $\mathbf{w}$ \emph{arbitrary} (but satisfying 1), by
using (\ref{16}) we obtain from (\ref{14}) that 
\begin{equation}
\overset{\cdot }{\overline{\left( \mathbf{F}^{T}\partial _{\mathbf{\dot{x}}}%
\mathcal{L}+\nabla \mathbf{\nu }^{T}\partial _{\mathbf{\dot{\nu}}}\mathcal{L}%
\right) }}-Div\left( \mathbb{P-}\left( \frac{1}{2}\rho _{0}\left\Vert 
\mathbf{\dot{x}}\right\Vert ^{2}+\chi \left( \mathbf{\nu ,\dot{\nu}}\right)
\right) \mathbf{I}\right) -\partial _{\mathbf{X}}\mathcal{L}=\mathbf{0.}
\label{23}
\end{equation}%
where $\mathbb{P}=e\mathbf{I}-\mathbf{F}^{T}\mathbf{P}-\nabla \mathbf{\nu }%
^{T}\underline{\ast }\mathcal{S}$ is the modified Eshelby tensor for
continua with substructure (see [4] for a similar result in a
non-conservative setting, where the elastic potential $e$ is substituted by
the free energy). $\mathbf{I}$ is the second-order unit tensor and the
product $\underline{\ast }$ is defined by $\left( \nabla \mathbf{\nu }^{T}%
\underline{\ast }\mathcal{S}\right) \mathbf{n\cdot u=}$ $\mathcal{S}\mathbf{%
n\cdot }\left( \nabla \mathbf{\nu }\right) \mathbf{u}$ for any pair of
vectors $\mathbf{n}$ and $\mathbf{u}$.

\textbf{Remark 5}. Let us assume as special choices that $\mathbf{\mathbf{f}}%
^{1}$ is such that $\mathbf{w=}\mathfrak{\dot{q}}\times \left( \mathbf{X}-%
\mathbf{X}_{0}\right) $ (with $\mathfrak{\dot{q}}$ a rigid rotational
velocity $\mathbf{X}_{0}$ a fixed point in space) and that $G=SO\left(
3\right) $, being $\mathfrak{\dot{q}}\times $ an element of its Lie algebra,
thus $\xi _{\mathcal{M}}\left( \mathbf{\nu }\right) =\mathcal{A}\mathfrak{%
\dot{q}}$. If the material is homogeneous, and we assume that $\mathbf{%
\mathbf{f}}^{1}$ and $G$ alone (in the form just defined) act on $\mathcal{L}
$, we have $skw\left( \mathbf{F}^{T}\partial _{\mathbf{F}}\mathcal{L}+\left(
\nabla \mathbf{\nu }\right) ^{T}\underline{\ast }\partial _{\nabla \mathbf{%
\nu }}\mathcal{L}\right) =0$.

\textbf{Remark 6}. The action of $\mathbf{\mathbf{f}}^{1}$can be interpreted
as a special virtual mutation of a possibly existing smooth distribution of
inhomogeneities throughout the body, in the sense of [8]. In other words, we
may say that (\ref{23}) is the balance of interactions arising when the body
mutates its inhomogeneous structure. This interpretation has been also
suggested in [9] in non-conservative setting.

\subsection{Hamilton equations}

Define $\mathbf{p}$ and $\mathbf{\mu }$, respectively the \emph{canonical
momentum} and the\ \emph{canonical substructural momentum},\ by%
\begin{equation}
\mathbf{p=}\partial _{\mathbf{\dot{x}}}\mathcal{L}\text{ \ \ , \ \ }\mathbf{%
\mu =}\partial _{\mathbf{\dot{\nu}}}\mathcal{L}\mathbf{.}  \label{24}
\end{equation}%
The \emph{Hamiltonian density} $\mathcal{H}$,%
\begin{equation}
\mathcal{H}\left( \mathbf{X,x,p,F,\nu ,\mu ,}\nabla \mathbf{\nu }\right) =%
\mathbf{p}\mathfrak{\cdot }\mathbf{\dot{x}}+\mathbf{\mu }\cdot \mathbf{\dot{%
\nu}}-\mathcal{L}\left( \mathbf{X,x,\dot{x},F,\nu ,\dot{\nu},}\nabla \mathbf{%
\nu }\right) .  \label{27}
\end{equation}%
has partial derivatives with respect to its entries; some of them are the
opposite of the corresponding derivatives of $\mathcal{L}$\ so that (\ref{3}%
), (\ref{4}) can be also written respectively as%
\begin{eqnarray}
\mathbf{\dot{p}} &\mathbf{=}&\mathbf{-}\partial _{\mathbf{x}}\mathcal{H}%
+Div\partial _{\mathbf{F}}\mathcal{H},  \notag \\
\mathbf{\dot{x}} &\mathbf{=}&\partial _{\mathbf{p}}\mathcal{H};  \label{28}
\end{eqnarray}%
\begin{eqnarray}
\mathbf{\dot{\mu}} &\mathbf{=}&-\partial _{\mathbf{\nu }}\mathcal{H}%
+Div\partial _{\nabla \mathbf{\nu }}\mathcal{H},  \notag \\
\mathbf{\dot{\nu}} &\mathbf{=}&\partial _{\mathbf{\mu }}\mathcal{H}.
\label{29}
\end{eqnarray}

\section{Canonical Poisson brackets in multifield theories}

We now consider a general boundary value problem where the following
boundary conditions are associated with (\ref{28}) and (\ref{29})%
\begin{equation}
\mathbf{x}\left( \mathbf{X}\right) =\mathbf{\bar{x}}\text{ \ \ \ \ \ on }%
\partial ^{\left( 1\right) }\mathcal{B}_{0},  \label{34}
\end{equation}%
\begin{equation}
\partial _{\mathbf{F}}\mathcal{H}\mathbf{n=t}\text{ \ \ \ \ \ \ \ \ \ \ on }%
\partial ^{\left( 2\right) }\mathcal{B}_{0},  \label{35}
\end{equation}%
\begin{equation}
\mathbf{\nu }\left( \mathbf{X}\right) =\mathbf{\bar{\nu}}\text{ \ \ \ \ \ \
on }\partial ^{\left( 1\right) }\mathcal{B}_{0},  \label{36}
\end{equation}%
\begin{equation}
\partial _{\nabla \mathbf{\nu }}\mathcal{H}\mathbf{n=}\mathfrak{t}\text{ \ \
\ \ \ \ \ \ \ \ \ on }\partial ^{\left( 2\right) }\mathcal{B}_{0};
\label{37}
\end{equation}%
$\mathbf{\bar{x}}$, $\mathbf{t}$, $\mathbf{\bar{\nu}}$\ and $\mathfrak{t}$\
are prescribed on the relevant parts of the boundary and $Cl\left( \partial 
\mathcal{B}_{0}\right) =Cl\left( \partial ^{\left( 1\right) }\mathcal{B}%
_{0}\cup \partial ^{\left( 2\right) }\mathcal{B}_{0}\right) $, with $%
\partial ^{\left( 1\right) }\mathcal{B}_{0}\cap \partial ^{\left( 2\right) }%
\mathcal{B}_{0}=\varnothing $, where $Cl$ indicates closure and $\mathbf{n}$
is the outward unit normal to $\partial \mathcal{B}_{0}$ at all points in
which it is well defined.

We assume that there exist two surface densities $\bar{U}\left( \mathbf{x}%
\right) $\ and $U\left( \mathbf{\nu }\right) $ such that%
\begin{equation}
t=\rho _{0}\partial _{\mathbf{x}}\bar{U}\text{ \ ,\ \ \ }\mathfrak{t}=\rho
_{0}\partial _{\mathbf{\nu }}U,  \label{38}
\end{equation}%
where $\bar{U}$\ and $U$ plays here the r\^{o}le of surface potentials.

Then the Hamiltonian $H$ of the whole body is given by%
\begin{equation}
H\left( \mathbf{X,x,p,\nu ,\mu }\right) =\int_{\mathcal{B}_{0}}\mathcal{H}%
\left( \mathbf{X,x,p,\nu ,\mu }\right) -\int_{\partial ^{\left( 2\right) }%
\mathcal{B}_{0}}\left( \bar{U}\left( \mathbf{x}\right) -U\left( \mathbf{\nu }%
\right) \right) .  \label{41}
\end{equation}%
Notice that we write $\mathcal{H}\left( \mathbf{X,x,p,\nu ,\mu }\right) $
instead of $\mathcal{H}\left( \mathbf{X,x,p,F,\nu ,\mu ,}\nabla \mathbf{\nu }%
\right) $ because below we consider directly variational derivatives.

\textbf{Theorem 2}. \emph{The canonical Hamilton equation}%
\begin{equation}
\dot{F}=\left\{ F,H\right\}   \label{42}
\end{equation}%
\emph{is equivalent to the Hamiltonian system of balance equations }(\ref{28}%
)-(\ref{29})\emph{\ for a continuum with substructure where F is any
functional of the type }$\int_{\mathcal{B}_{0}}f\left( \mathbf{X,x,p,\nu
,\mu }\right) $\emph{, with f a sufficiently smooth scalar density, and the
Poisson bracket }$\left\{ \cdot ,\cdot \right\} $\emph{\ for a complex
material is given by }%
\begin{eqnarray}
\left\{ F,H\right\}  &=&\int_{\mathcal{B}_{0}}\left( \frac{\delta f}{\delta 
\mathbf{x}}\cdot \frac{\delta \mathcal{H}}{\delta \mathbf{p}}-\frac{\delta 
\mathcal{H}}{\delta \mathbf{x}}\cdot \frac{\delta f}{\delta \mathbf{p}}%
\right) +  \notag \\
&&+\int_{\partial ^{\left( 2\right) }\mathcal{B}_{0}}\left( \frac{\delta f}{%
\delta \mathbf{x}}\cdot \frac{\delta \mathcal{H}}{\delta \mathbf{p}}%
\left\vert _{\partial ^{\left( 2\right) }\mathcal{B}_{0}}\right. -\frac{%
\delta \mathcal{H}}{\delta \mathbf{x}}\cdot \frac{\delta f}{\delta \mathbf{p}%
}\left\vert _{\partial ^{\left( 2\right) }\mathcal{B}_{0}}\right. \right) + 
\notag \\
&&+\int_{\mathcal{B}_{0}}\left( \frac{\delta f}{\delta \mathbf{\nu }}\cdot 
\frac{\delta \mathcal{H}}{\delta \mathbf{\mu }}-\frac{\delta \mathcal{H}}{%
\delta \mathbf{\mu }}\cdot \frac{\delta f}{\delta \mathbf{\nu }}\right) + 
\notag \\
&&+\int_{\partial ^{\left( 2\right) }\mathcal{B}_{0}}\left( \frac{\delta f}{%
\delta \mathbf{\nu }}\cdot \frac{\delta \mathcal{H}}{\delta \mathbf{\mu }}%
\left\vert _{\partial ^{\left( 2\right) }\mathcal{B}_{0}}\right. -\frac{%
\delta \mathcal{H}}{\delta \mathbf{\nu }}\cdot \frac{\delta f}{\delta 
\mathbf{\mu }}\left\vert _{\partial ^{\left( 2\right) }\mathcal{B}%
_{0}}\right. \right) ,  \label{40}
\end{eqnarray}%
\emph{where the variational derivative} $\frac{\delta \mathcal{H}}{\delta 
\mathbf{x}}$ \emph{is obtained fixing }$\mathbf{p}$\emph{\ and allowing }$%
\mathbf{x}$\emph{\ to vary\footnote{%
See relevant remarks in [10].}; an analogous meaning is valid for the
variational derivative with respect to the order parameter.}

\ \ \ \ 

The proof can be developed by direct calculation. Clearly, $\left\{ \cdot
,\cdot \right\} $\ is \emph{bilinear} and \emph{skew-symmetric}, and one can
check easily that it \emph{satisfies the Jacobi's identity}. We note that%
\begin{eqnarray}
\left\{ F,H\right\}  &=&\int_{\mathcal{B}_{0}}\left( \frac{\delta f}{\delta 
\mathbf{x}}\cdot \frac{\partial \mathcal{H}}{\partial \mathbf{p}}-\frac{%
\delta f}{\delta \mathbf{p}}\cdot \left( \partial _{\mathbf{x}}\mathcal{H}%
-Div\partial _{\mathbf{F}}\mathcal{H}\right) \right) +  \notag \\
&&+\int_{\partial ^{\left( 2\right) }\mathcal{B}_{0}}\left( \frac{\delta f}{%
\delta \mathbf{x}}\cdot \frac{\partial \mathcal{H}}{\partial \mathbf{p}}%
\left\vert _{\partial ^{\left( 2\right) }\mathcal{B}_{0}}\right. -\frac{%
\delta f}{\delta \mathbf{p}}\cdot \left( \partial _{\mathbf{x}}\bar{U}%
\mathbf{-\partial _{\mathbf{F}}}\mathcal{H}\mathbf{n}\right) \left\vert
_{\partial ^{\left( 2\right) }\mathcal{B}_{0}}\right. \right) +  \notag \\
&&+\int_{\mathcal{B}_{0}}\left( \frac{\delta f}{\delta \mathbf{\nu }}\cdot 
\frac{\partial \mathcal{H}}{\partial \mathbf{\mu }}-\frac{\delta f}{\delta 
\mathbf{\mu }}\cdot \left( \partial _{\mathbf{\nu }}\mathcal{H}-Div\partial
_{\nabla \mathbf{\nu }}\mathcal{H}\right) \right) +  \notag \\
&&+\int_{\partial ^{\left( 2\right) }\mathcal{B}_{0}}\left( \frac{\delta f}{%
\delta \mathbf{\nu }}\cdot \frac{\partial \mathcal{H}}{\partial \mathbf{\mu }%
}\left\vert _{\partial ^{\left( 2\right) }\mathcal{B}_{0}}\right. -\frac{%
\delta f}{\delta \mathbf{\mu }}\cdot \left( \mathbf{\partial }_{\mathbf{\nu }%
}U\mathbf{-\partial _{\nabla \mathbf{\nu }}}\mathcal{H}\mathbf{n}\right)
\left\vert _{\partial ^{\left( 2\right) }\mathcal{B}_{0}}\right. \right) ,
\label{43}
\end{eqnarray}%
and, in terms of functional partial derivatives,%
\begin{eqnarray}
\dot{F} &=&\int_{\mathcal{B}_{0}}\left( \frac{\delta f}{\delta \mathbf{x}}%
\cdot \mathbf{\dot{x}}+\frac{\delta f}{\delta \mathbf{p}}\cdot \mathbf{\dot{p%
}+}\frac{\delta f}{\delta \mathbf{\nu }}\cdot \mathbf{\dot{\nu}+}\frac{%
\delta f}{\delta \mathbf{\mu }}\cdot \mathbf{\dot{\mu}}\right) +  \notag \\
&&+\int_{\partial ^{\left( 2\right) }\mathcal{B}_{0}}\frac{\delta f}{\delta 
\mathbf{x}}\cdot \mathbf{\dot{x}}\left\vert _{\partial ^{\left( 2\right) }%
\mathcal{B}_{0}}\right. \mathbf{+}\int_{\partial ^{\left( 2\right) }\mathcal{%
B}_{0}}\frac{\delta f}{\delta \mathbf{p}}\cdot \mathbf{\dot{p}}\left\vert
_{\partial ^{\left( 2\right) }\mathcal{B}_{0}}\right. .  \label{44}
\end{eqnarray}%
By identifying analogous terms in (\ref{43}) and (\ref{44}), we obtain both
the Hamiltonian system (\ref{28})-(\ref{29}) and the boundary conditions (%
\ref{34})-(\ref{37}).

When we put $F=H$, (\ref{42}) coincides with the equation of \emph{%
conservation of energy}. We have, in fact,%
\begin{equation}
\dot{H}=\left\{ H,H\right\} =0.  \label{45}
\end{equation}

Geometrical properties of the Poisson brackets for direct models of rods,
plates and complex fluids have been discussed in [10], [11].

\section{A formal approach toward an Hamilton-Jacoby theory with gradient
effects}

Let $h$ be a smooth diffeomorphism%
\begin{equation}
h:\left( \mathbf{X,x,p,F,\nu ,\mu ,}\nabla \mathbf{\nu }\right) \longmapsto
\left( \mathbf{X,x}_{\ast }\mathbf{,p}_{\ast }\mathbf{,F}_{\ast }\mathbf{%
,\nu }_{\ast }\mathbf{,\mu }_{\ast }\mathbf{,}\nabla \mathbf{\nu }_{\ast
}\right) .  \label{46}
\end{equation}%
The transformation $h$ generates a new Hamiltonian density%
\begin{equation}
\mathcal{H}_{\ast }\left( \mathbf{X,x}_{\ast }\mathbf{,p}_{\ast }\mathbf{,F}%
_{\ast }\mathbf{,\nu }_{\ast }\mathbf{,\mu }_{\ast }\mathbf{,}\nabla \mathbf{%
\nu }_{\ast }\right) ,  \label{47}
\end{equation}%
with corresponding Lagrangian density%
\begin{equation}
\mathcal{L}_{\ast }=\mathbf{p}_{\ast }\cdot \mathbf{\dot{x}}_{\ast }+\mathbf{%
\mu }_{\ast }\cdot \mathbf{\dot{\nu}}_{\ast }-\mathcal{H}_{\ast }.
\label{48}
\end{equation}%
If $h$ were such that $\mathcal{H}_{\ast }=0$, then an immediate integration
of the system (\ref{28}), (\ref{29}) could be achieved. To this aim we
choose $h$ to be such that the integral of the difference $\mathcal{L}-%
\mathcal{L}_{\ast }$ between two instants, say $t_{1}$ and $t_{2}$, be equal
to the time derivative of a generating function $S$ of the type $S=S\left( t,%
\mathbf{X,x,p}_{\ast }\mathbf{,\nu ,\mu }_{\ast }\right) $, i.e.%
\begin{equation}
\int_{t_{1}}^{t_{2}}\mathcal{L}-\mathcal{L}_{\ast }=S\left\vert
_{t=t_{2}}\right. -S\left\vert _{t=t_{1}}\right. .  \label{49}
\end{equation}%
Then, from (\ref{49}) we would have%
\begin{equation*}
\left( \mathbf{p}\cdot \mathbf{\dot{x}}+\mathbf{\mu }\cdot \mathbf{\dot{\nu}}%
-\mathcal{H}\right) -\left( \mathbf{p}_{\ast }\cdot \mathbf{\dot{x}}_{\ast }+%
\mathbf{\mu }_{\ast }\cdot \mathbf{\dot{\nu}}_{\ast }-\mathcal{H}_{\ast
}\right) =
\end{equation*}%
\begin{equation}
=\dot{S}=\partial _{t}S+\partial _{\mathbf{x}}S\cdot \mathbf{\dot{x}%
+\partial }_{\mathbf{p}_{\ast }}S\cdot \mathbf{\dot{p}}_{\ast }+\partial _{%
\mathbf{\nu }}S\cdot \mathbf{\dot{\nu}+\partial }_{\mathbf{\mu }_{\ast
}}S\cdot \mathbf{\dot{\mu}}_{\ast },  \label{50}
\end{equation}%
and hence%
\begin{equation}
\mathbf{p=}\partial _{\mathbf{x}}S\text{ \ \ , \ \ }\mathbf{\mu =}\partial _{%
\mathbf{\nu }}S\text{,}  \label{51}
\end{equation}%
\begin{equation}
\mathbf{x}_{\ast }-\mathbf{x}_{0}=\mathbf{\partial }_{\mathbf{p}_{\ast }}S%
\text{ \ \ , \ \ }\mathbf{\nu }=\mathbf{\partial }_{\mathbf{\mu }_{\ast }}S,
\label{52}
\end{equation}%
\begin{equation}
\partial _{t}S+\mathcal{H}=\mathcal{H}_{\ast }.  \label{53}
\end{equation}

To obtain (\ref{52}) one makes use of the fact that $\delta
\int_{t_{1}}^{t_{2}}\left( \mathbf{p\cdot }\left( \mathbf{x}-\mathbf{x}%
_{0}\right) +\mathbf{\mu }\cdot \mathbf{\nu }\right) =0$ for variations
vanishing at $t_{1}$ and $t_{2}$ (in the sense that $\delta \left( \mathbf{%
\mu }\cdot \mathbf{\nu }\right) \left\vert _{t=t_{1},t_{2}}=0\right. $ and $%
\delta \left( \mathbf{p\cdot }\left( \mathbf{x}-\mathbf{x}_{0}\right)
\right) \left\vert _{t=t_{1},t_{2}}=0\right. $) so that $\mathbf{p}\cdot 
\mathbf{\dot{x}=\dot{p}\cdot }\left( \mathbf{x}-\mathbf{x}_{0}\right) $ and $%
\mathbf{\mu }\cdot \mathbf{\dot{\nu}=\dot{\mu}}\cdot \mathbf{\nu }$.

A necessary and sufficient condition to assure that $\mathcal{H}_{\ast }=0$\
is%
\begin{equation}
\partial _{t}S+\mathcal{H}\left( \mathbf{X,x,\partial _{\mathbf{x}}}S\mathbf{%
,F,\nu ,\partial _{\mathbf{\nu }}}S\mathbf{,}\nabla \mathbf{\nu }\right) =0,
\label{54}
\end{equation}%
which is a Hamiltonian-Jacobi like equation. Since $\mathcal{H}_{\ast }=0$, $%
\mathbf{p}_{\ast }\ $and $\mathbf{\mu }_{\ast }$ are constant in time, the
time derivative of $S$ reduces to%
\begin{equation}
\dot{S}=\partial _{t}S+\partial _{\mathbf{x}}S\cdot \mathbf{\dot{x}+}%
\partial _{\mathbf{\nu }}S\cdot \mathbf{\dot{\nu}=-}\mathcal{H}\mathbf{+p}%
\cdot \mathbf{\dot{x}}+\mathbf{\mu }\cdot \mathbf{\dot{\nu}=}\mathcal{L}.
\label{55}
\end{equation}

The relation (\ref{55}) allows us to determine $S$ to within a constant,
namely%
\begin{equation}
S=\int \mathcal{L}dt+const.  \label{56}
\end{equation}

\section{The spatial form}

Circumstances in which the notion of reference placement is wanting, as in
the case of fluids or granular flows, render the choice of a material or
spatial representation not matter of form only (see, e.g., [12] for standard
bodies). Here, having in mind the study of complex fluids, we provide a
spatial variational derivation of the balance equations \emph{free of any
concept of reference place or paragon setting and without even formal
recourse to an inverse motion}. So, in the present section $\mathbf{x}\in 
\mathcal{B}$ is just a point in space. The notation $\mathbf{u}\left( 
\mathbf{x},t\right) $ is used for the velocity field over $\mathcal{B}$. The
order parameter is now $\mathbf{\nu }\left( \mathbf{x,}t\right) $ and we
indicate with $\mathbf{\upsilon }\left( \mathbf{x,}t\right) $ its actual
rate. The symmetric tensor $\mathbf{g}$ is the \emph{spatial metric}
characterizing the present state of the body; it plays a prominent r\^{o}le
because in this case the counterpart of (\ref{2}) of the Lagrangian density
is of the form%
\begin{equation}
\mathcal{L}\left( \mathbf{x,u,g,\nu ,\upsilon ,}grad\mathbf{\nu }\right) =%
\frac{1}{2}\rho \left\Vert \mathbf{v}\right\Vert ^{2}+\rho \chi \left( 
\mathbf{\nu ,\upsilon }\right) -\rho e\left( \mathbf{g,\nu ,}grad\mathbf{\nu 
}\right) -\rho w\left( \mathbf{x,\nu }\right) ,  \label{57}
\end{equation}%
with some slight abuse of notation. We then find balance equations as
conditions verifying the relation%
\begin{equation}
\hat{\delta}\left( \int_{0}^{\bar{t}}d\tau \int_{\mathcal{B}}\mathcal{L}%
\left( \mathbf{x,u,g,\nu ,\upsilon ,}grad\mathbf{\nu }\right) \right) =0,
\label{58}
\end{equation}%
where $\hat{\delta}$ denotes the total variation.

To define the variation of the relevant fields, we make use of $\mathbf{f}%
^{2}$ introduced at point 2 of Section 1 and identify $\delta \mathbf{x}$
with $\mathbf{v}$. We consider a special (though wide) subclass of possible
vector fields $\mathbf{x\longmapsto v}\left( \mathbf{x}\right) $
characterized by the circumstance that they are \emph{purely} deformative;
in other words, we choose $\mathbf{v}$ such that $skewgrad\mathbf{v=0}$.

We then define%
\begin{equation}
\hat{\delta}\mathbf{g=}\frac{d}{ds_{2}}\mathbf{f}^{2\ast }\mathbf{g}%
\left\vert _{s_{2}=0}\right. =L_{\mathbf{v}}\mathbf{g}=2symgrad\mathbf{v}%
=2grad\mathbf{v},  \label{59}
\end{equation}%
where $\mathbf{f}^{2\ast }$ means pull back and $L_{\mathbf{v}}$ is thus the
autonomous Lie derivative following the flow $\mathbf{v}$. In analogous way,
we put%
\begin{equation}
\hat{\delta}\mathbf{\nu }=\delta \mathbf{\nu }+\left( grad\mathbf{\nu }%
\right) \mathbf{v},  \label{60}
\end{equation}%
\begin{equation}
grad\hat{\delta}\mathbf{\nu }=grad\hat{\delta}\mathbf{\nu }+\left( grad%
\mathbf{\nu }\right) grad\mathbf{v}.  \label{61}
\end{equation}

As an intermediate step we notice that%
\begin{equation*}
\hat{\delta}\int_{\mathcal{B}}e\left( \mathbf{g,\nu ,}grad\mathbf{\nu }%
\right) =\int_{\mathcal{B}}\hat{\delta}e=
\end{equation*}%
\begin{equation}
=\int_{\mathcal{B}}\left( 2\partial _{\mathbf{g}}e\cdot grad\mathbf{%
v+\partial }_{\mathbf{\nu }}e\cdot \hat{\delta}\mathbf{\nu }+\partial _{grad%
\mathbf{\nu }}e\cdot \left( grad\hat{\delta}\mathbf{\nu }+\left( grad\mathbf{%
\nu }\right) grad\mathbf{v}\right) \right) .  \label{62}
\end{equation}

By developing the variation of (\ref{58}), making use of (\ref{59})-(\ref{62}%
) and Gauss theorem, we recognize that appropriate balances in the bulk are%
\begin{equation}
\overset{\cdot }{\overline{\partial _{\mathbf{u}}\mathcal{L}}}-\partial _{%
\mathbf{x}}\mathcal{L}+div\left( 2\partial _{\mathbf{g}}\mathcal{L}-\left(
grad\mathbf{\nu }\right) ^{T}\partial _{grad\mathbf{\nu }}\mathcal{L}\right)
=0,  \label{63}
\end{equation}%
\begin{equation}
\overset{\cdot }{\overline{\partial _{\mathbf{\upsilon }}\mathcal{L}}}%
-\partial _{\mathbf{\nu }}\mathcal{L}+div\left( \partial _{grad\mathbf{\nu }}%
\mathcal{L}\right) =0.  \label{64}
\end{equation}%
Cauchy stress $\mathbf{T}$\ is then given by%
\begin{equation}
\mathbf{T=-}2\partial _{\mathbf{g}}\mathcal{L}-\left( grad\mathbf{\nu }%
\right) ^{T}\mathcal{S}_{a},  \label{65}
\end{equation}%
where the \emph{actual} microstress $\mathcal{S}_{a}$\ is defined by%
\begin{equation}
\mathcal{S}_{a}=-\partial _{grad\mathbf{\nu }}\mathcal{L}.  \label{66}
\end{equation}

In the case of simple bodies, (\ref{65}) reduces to the well known
Doyle-Ericksen formula.

\textbf{Remark 7}. A requirement of invariance of $e$ under the action of $%
SO\left( 3\right) $ implies that%
\begin{equation}
skew\left( 2\partial _{\mathbf{g}}\mathcal{L}\right) =\mathsf{e}\left( 
\mathcal{A}^{T}\mathbf{z}_{a}+\left( grad\mathcal{A}^{T}\right) \mathcal{S}%
_{a}\right) ,  \label{67}
\end{equation}%
where $\mathbf{z}_{a}=-\rho \partial _{\mathbf{\nu }}e$ is the actual
self-force and $\mathsf{e}$ Ricci's alternating tensor.

\subsection{Spatial Hamilton equations}

To find appropriate spatial Hamilton equations, we follow the pattern of
Section 1.1. To this end we define spatial canonical standard and
substructural momenta ($\mathbf{\bar{p}}$ and $\mathbf{\bar{\mu}}$
respectively) through%
\begin{equation}
\mathbf{\bar{p}=}\partial _{\mathbf{u}}\mathcal{L}\text{ \ \ , \ \ }\mathbf{%
\bar{\mu}=}\partial _{\mathbf{\upsilon }}\mathcal{L}\mathbf{.}  \label{68}
\end{equation}

Consequently, the spatial Hamiltonian density is given by%
\begin{equation}
\mathcal{H}\left( \mathbf{x,\bar{p},g,\nu ,\bar{\mu},}grad\mathbf{\nu }%
\right) =\mathbf{\bar{p}\cdot u+\bar{\mu}\cdot \upsilon -}\mathcal{L}\left( 
\mathbf{x,u,g,\nu ,\upsilon ,}grad\mathbf{\nu }\right)  \label{69}
\end{equation}%
(with some slight abuse of notation) and has partial derivatives with
respect to its entries. By evaluating the variation of $\mathcal{H}$, taking
into account (\ref{59}) and (\ref{61}), and comparing the result with the
variation of $\mathcal{L}$, after making use of the balances (\ref{63}) and (%
\ref{64}), we obtain the spatial form of the Hamilton equations:%
\begin{eqnarray}
\overset{\cdot }{\mathbf{\bar{p}}} &=&\mathbf{-}\partial _{\mathbf{x}}%
\mathcal{H}+div\left( 2\partial _{\mathbf{g}}\mathcal{H}-\left( grad\mathbf{%
\nu }\right) ^{T}\partial _{grad\mathbf{\nu }}\mathcal{H}\right) ,  \notag \\
\mathbf{u} &=&\partial _{\mathbf{\bar{p}}}\mathcal{H};  \label{70}
\end{eqnarray}%
\begin{eqnarray}
\overset{\cdot }{\mathbf{\bar{\mu}}} &=&\mathbf{-}\partial _{\mathbf{\nu }}%
\mathcal{H}+div\left( \partial _{grad\mathbf{\nu }}\mathcal{H}\right) , 
\notag \\
\mathbf{\upsilon } &=&\partial _{\mathbf{\bar{\mu}}}\mathcal{H}.  \label{71}
\end{eqnarray}

\subsection{Spatial Hamilton-Jacobi form}

We may obtain the spatial counterpart of (\ref{54}) by considering a smooth
diffeomorphism%
\begin{equation}
\bar{h}:\left( \mathbf{x,\bar{p},g,\nu ,\bar{\mu},}grad\mathbf{\nu }\right)
\longmapsto \left( \mathbf{x}_{\ast }\mathbf{,\bar{p}}_{\ast }\mathbf{,g}%
_{\ast }\mathbf{,\nu }_{\ast }\mathbf{,\bar{\mu}}_{\ast }\mathbf{,}grad%
\mathbf{\nu }_{\ast }\right) ,  \label{72}
\end{equation}%
which generates a new Hamiltonian density%
\begin{equation}
\mathcal{H}_{\ast }\left( \mathbf{x}_{\ast }\mathbf{,\bar{p}}_{\ast }\mathbf{%
,g}_{\ast }\mathbf{,\nu }_{\ast }\mathbf{,\bar{\mu}}_{\ast }\mathbf{,}grad%
\mathbf{\nu }_{\ast }\right) .  \label{73}
\end{equation}

Now, we may use a generating function $S=S\left( t,\mathbf{x,\mathbf{\bar{p}}%
_{\ast },\nu ,\bar{\mu}}_{\ast }\right) $, and, following the same procedure
of Section 3, we find that a necessary and sufficient condition to assure
that $\mathcal{H}_{\ast }=0$ is%
\begin{equation}
\partial _{t}S+\mathcal{H}\left( \mathbf{x,\partial }_{\mathbf{x}}S\mathbf{%
,g,\nu ,\mathbf{\partial }_{\mathbf{\nu }}}S\mathbf{,}grad\mathbf{\nu }%
\right) =0.  \label{74}
\end{equation}

\subsection{A spatial form of Poisson brackets}

For the spatial Hamiltonian in equations (\ref{70})-(\ref{71}), taking into
account (\ref{59})-(\ref{61}), we define a new variational derivative $%
\overline{\frac{\partial \mathcal{H}}{\partial \mathbf{x}}}$ through the
relation%
\begin{equation}
\overline{\frac{\delta \mathcal{H}}{\delta \mathbf{x}}}\left( \mathbf{x,%
\mathbf{\bar{p}},\nu ,\bar{\mu}}\right) \cdot \mathbf{v=}\left( \mathbf{-}%
\partial _{\mathbf{x}}\mathcal{H}+div\left( 2\partial _{\mathbf{g}}\mathcal{H%
}-\left( grad\mathbf{\nu }\right) ^{T}\partial _{grad\mathbf{\nu }}\mathcal{H%
}\right) \right) \cdot \mathbf{v,}  \label{75}
\end{equation}%
holding $\mathbf{\bar{p}}$ fixed and allowing $\mathbf{x}$ to vary, for any $%
\mathbf{v}$ of the kind used in (\ref{59})-(\ref{61}).

Consider a boundary value problem of the type%
\begin{equation}
\left( 2\partial _{\mathbf{g}}\mathcal{H}-\left( grad\mathbf{\nu }\right)
^{T}\partial _{grad\mathbf{\nu }}\mathcal{H}\right) \mathbf{n=\partial }_{%
\mathbf{x}}\bar{u}\left( \mathbf{x}\right) \text{ , }\left( \partial _{grad%
\mathbf{\nu }}\mathcal{H}\right) \mathbf{n=\partial }_{\mathbf{\nu }}u\left( 
\mathbf{\nu }\right) \text{ , on }\partial \mathcal{B},  \label{76}
\end{equation}%
(where $\bar{u}\left( \mathbf{x}\right) $\ and $u\left( \mathbf{\nu }\right) 
$\ are the counterparts of the surface potentials $\bar{U}\left( \mathbf{x}%
\right) $ and $U\left( \mathbf{\nu }\right) $).

The total Hamiltonian is now given by $H\left( \mathbf{x,\mathbf{\bar{p}}%
,\nu ,\bar{\mu}}\right) =\int_{\mathcal{B}}\mathcal{H}$ (with some slight
abuse of notation) and we list only the entries $\left( \mathbf{x,\mathbf{%
\bar{p}},\nu ,\bar{\mu}}\right) $ because we consider the variational
derivative (\ref{70}) below. We consider also arbitrary functionals $F$ of
the type $\int_{\mathcal{B}}f\left( \mathbf{x,\mathbf{\bar{p}},\nu ,\bar{\mu}%
}\right) $, with $f$ a sufficiently smooth scalar density.

\textbf{Theorem 3}. \emph{The canonical Hamilton equation}%
\begin{equation}
\dot{F}=\left\{ F,H\right\} _{a}  \label{77}
\end{equation}%
\emph{is equivalent to the Hamiltonian system of balance equations }(\ref{70}%
)-(\ref{71})\emph{\ with }%
\begin{eqnarray}
\left\{ F,H\right\} _{a} &=&\int_{\mathcal{B}}\left( \overline{\frac{\delta f%
}{\delta \mathbf{x}}}\cdot \frac{\delta \mathcal{H}}{\delta \mathbf{p}}-%
\overline{\frac{\delta \mathcal{H}}{\delta \mathbf{x}}}\cdot \frac{\delta f}{%
\delta \mathbf{p}}\right) +  \notag \\
&&+\int_{\partial \mathcal{B}}\left( \overline{\frac{\delta f}{\delta 
\mathbf{x}}}\cdot \frac{\delta \mathcal{H}}{\delta \mathbf{p}}\left\vert
_{\partial \mathcal{B}}\right. -\overline{\frac{\delta \mathcal{H}}{\delta 
\mathbf{x}}}\cdot \frac{\delta f}{\delta \mathbf{p}}\left\vert _{\partial 
\mathcal{B}}\right. \right) +  \notag \\
&&+\int_{\mathcal{B}}\left( \frac{\delta f}{\delta \mathbf{\nu }}\cdot \frac{%
\delta \mathcal{H}}{\delta \mathbf{\mu }}-\frac{\delta \mathcal{H}}{\delta 
\mathbf{\mu }}\cdot \frac{\delta f}{\delta \mathbf{\nu }}\right) +  \notag \\
&&+\int_{\partial \mathcal{B}}\left( \frac{\delta f}{\delta \mathbf{\nu }}%
\cdot \frac{\delta \mathcal{H}}{\delta \mathbf{\mu }}\left\vert _{\partial 
\mathcal{B}}\right. -\frac{\delta \mathcal{H}}{\delta \mathbf{\nu }}\cdot 
\frac{\delta f}{\delta \mathbf{\mu }}\left\vert _{\partial \mathcal{B}%
}\right. \right) ,  \label{78}
\end{eqnarray}%
\emph{where }$\left\{ \cdot ,\cdot \right\} _{a}$ \emph{is bilinear,
skew-symmetric and satisfies Jacobi's identity}.

\section{Final remarks}

To illustrate possible uses of Theorem 2, we list below some special cases.
Analogous results accrue from Theorem 3.

\textbf{Remark 8}. If we choose $f=\mathbf{p\cdot v}$, with $\mathbf{v}$ an
arbitrary vector, equation (\ref{28}$_{a}$) and the boundary condition (\ref%
{35}) follow immediately from (\ref{42}).

\textbf{Remark 9}. Let $f=\mathbf{\mu \cdot }\xi _{\mathcal{M}}\left( 
\mathbf{\nu }\right) $, then from (\ref{42}) we get (\ref{29}$_{a}$) and the
boundary condition (\ref{37}).

\textbf{Remark 10}. Let $f$ be of the form%
\begin{equation}
f=\mathbf{p\cdot }\left( \mathbf{\dot{q}}\times \left( \mathbf{x}-\mathbf{x}%
_{0}\right) \right) +\mathbf{\mu \cdot }\mathcal{A}\mathbf{\dot{q},}
\label{81}
\end{equation}%
with $\mathbf{\dot{q}}$\ arbitrary as in previous sections. Consider also,
for the sake of simplicity, absence of external bulk interactions (the ones
accounted for $w\left( \mathbf{x,\nu }\right) $). By using (\ref{28}) and (%
\ref{29}), we obtain from (\ref{42})%
\begin{equation}
\mathsf{e}\left( \partial _{\mathbf{F}}\mathcal{H}\mathbf{F}^{T}\right) =%
\mathcal{A}^{T}\partial _{\mathbf{\nu }}\mathcal{H}+\left( \nabla \mathcal{A}%
^{T}\right) ^{t}\partial _{\nabla \mathbf{\nu }}\mathcal{H}.  \label{82}
\end{equation}

These remarks are the Hamiltonian counterparts of Remarks 1, 2, 3. Of
course, Poisson parentheses not only allow one to write in a concise form
balance equations, but generate articulated geometric structures over the
infinite-dimensional manifold of mappings showing placements and order
parameters, and properties of these structures depend also on the geometric
properties of $\mathcal{M}$.

\ \ \ \ \ \ 

\textbf{Acknowledgement}. This paper is the extended version of the first
part of a communication of PMM delivered at the Symposium honoring the
memory of Clifford Ambrose Truesdell III, within the 14$^{th}$ US National
Congress of Theoretical and Applied Mechanics, Blacksburg, June 2002. PMM
acknowledges gratefully the support of the Department of Mathematics of the
University of Kentucky (through C.-S. Man). We also thank Reuven Segev for
valuable discussions. The support of the Italian National Group of
Mathematical Physics (INDAM-GNFM) is acknowledged.

\section{References}

\begin{enumerate}
\item Capriz, G., \emph{Continua with microstructure}, Springer Verlag,
Berlin, 1989.

\item Capriz, G. (2000), Continua with substructure, \emph{Phys. Mesomech.}, 
\textbf{3}, 5-14, 37-50.

\item Capriz, G. and Mariano, P. M. (2002), Balance at a junction among
coherent interfaces in materials with substructure, in \emph{Advances in
multifield theories of materials with substructure}, G. Capriz and P. M.
Mariano edts, Birkhauser, in press.

\item Mariano, P. M. (2001), Multifield theories in mechanics of solids, 
\emph{Adv. Appl. Mech.}, \textbf{38}, 1-93.

\item Segev, R. (1994), A geometrical framework for the statics of materials
with microstructure, \emph{Math. Mod. Meth. Appl. Sci.}, \textbf{4}, 871-897.

\item Marsden, J.\ E. and Hughes, T. J. R., \emph{Mathematical foundations
of elasticity}, Prentice-Hall, Englewood Cliffs, New Jersey, 1983.

\item Abraham, R., Marsden, J. E., \emph{Foundations of mechanics},
Benjamin/Cummings Publishing Company, 1978.

\item Noll, W. (1967), Materially uniform simple bodies with
inhomogeneities, \emph{Arch. Rational Mech. Anal.}, \textbf{27}, 1-32.

\item Epstein, M. (2002), The Eshelby tensor and the theory of continuous
distributions of inhomogeneities, \emph{Mech. Res. Comm.}, \textbf{29},
501-506.

\item Simo, J. C., Marsden, J. E. and Krishnaprasad, P. S. (1988), The
Hamiltonian structure of nonlinear elasticity: The material and convective
representation of solids, rods and plates, \emph{Arch. Rational Mech. Anal. }%
\textbf{104}, 125-183.

\item Cendra, H., Marsden, J. E. and Ratiu, T. S. (2002), Cocycles,
compatibility and Poisson brackets for complex fluids, in \emph{Advances in
multifield theories of materials with substructure}, G. Capriz and P. M.
Mariano edts, Birkhauser, in press.

\item Capriz, G. (1984), Spatial variational principles in continuum
mechanics, \emph{Arch. Rational Mech. Anal.}, \textbf{85}, 99-109.
\end{enumerate}

\end{document}